\documentclass[sigconf]{acmart}
\usepackage{hyperref}
\usepackage{stfloats}
\AtBeginDocument{%
  }

\copyrightyear{2026}
\acmYear{2026}
\setcopyright{cc}
\setcctype{by}
\acmConference[L@S '26]{Proceedings of the Thirteenth ACM Conference on Learning @ Scale}{June 29-July 03, 2026}{Seoul, Republic of Korea}
\acmBooktitle{Proceedings of the Thirteenth ACM Conference on Learning @ Scale (L@S '26), June 29-July 03, 2026, Seoul, Republic of Korea}
\acmDOI{10.1145/3774398.3811601}
\acmISBN{979-8-4007-2293-6/2026/06}
\begin{document}

\title{Coasting Through Class: Learning Opportunity Loss from Practice Avoidance During Individual Seatwork}

% \author{Anonymized}
\author{Ashish Gurung}
\affiliation{%
  \institution{Carnegie Mellon University}
  \city{Pittsburgh, PA}
  \country{USA}}
\email{agurung@cmu.edu}

\author{Jordan Gutterman}
\affiliation{%
  \institution{Carnegie Mellon University}
  \city{Pittsburgh, PA}
  \country{USA}}
\email{jgutterm@andrew.cmu.edu}

\author{Danielle R. Thomas}
\affiliation{%
  \institution{Carnegie Mellon University}
  \city{Pittsburgh, PA}
  \country{USA}}
\email{dchine@andrew.cmu.edu}

\author{Mingyu Feng}
\affiliation{%
  \institution{WestEd}
  \city{San Francisco, CA}
  \country{USA}}
\email{mingyufeng@gmail.com}

\author{Vincent Aleven}
\affiliation{%
  \institution{Carnegie Mellon University}
  \city{Pittsburgh, PA}
  \country{USA}}
\email{aleven@cs.cmu.edu}

\author{Kenneth R. Koedinger}
\affiliation{%
  \institution{Carnegie Mellon University}
  \city{Pittsburgh, PA}
  \country{USA}}
\email{koedinger@cmu.edu}

\renewcommand{\shortauthors}{Gurung et al.}

\begin{abstract}
Measures of disengagement provide insights into unproductive use of learning opportunities. Although measures of active disengagement, such as gaming the system and mind-wandering, are well studied, loss of practice time due to outright task avoidance remains relatively understudied. The current study addresses this gap by extending existing within-task measures (idle time) with two new session-level measures (delayed start and early stop) to capture loss of practice time due to task avoidance. We characterize the combined lost time as \textit{coasted time} and the associated behavior as \textit{coasting behavior}. Using ASSISTments logs (N = 1,425), we find that students dedicate only 40\% of available classwork time to math practice and coast through the remaining 60\%. Of the coasted time, 36\% resulted from delayed starts, 2\% from mid-practice idling, and 62\% from stopping early. Delayed start and early stop showed moderate temporal stability (G = 0.73 and 0.71, respectively), suggesting that coasting is a consistent behavioral pattern. Even after excluding early stops attributable to assignment completion (i.e., early stop = 0), coasted time remained substantial at 32\%. While we observe significant differences in coasting by gender and IEP status, we do not observe them by other demographic factors or school locale. Critically, students who continued working beyond the first assignment completion (\textit{``extra effort''}) performed significantly better on standardized tests. For research, coasting offers a new lens on opportunity loss by combining session-level disengagement with within-task disengagement. For practitioners, our results highlight the need for platform affordances that support sustained engagement and more productive use of available practice time.
\end{abstract}





\maketitle

\section{Introduction}

Effective use of learning opportunities is a fundamental determinant of learning. Digital learning platforms (DLPs) are designed to support effective use of learning opportunities through automated, adaptive support. However, cognitive or behavioral disengagement during practice can lead to suboptimal or unproductive use of these opportunities. Consequently, significant research effort has been dedicated towards leveraging learner log data in the detection and mitigation of learner disengagement, including gaming the system~\cite{baker2008students}, mind-wandering~\cite{d2017zone}, (low) effort~\cite{shih2011response,gurung2021examining}, procrastination~\cite{steel2007nature}, and off-task behavior~\cite{baker2004off}. Building on these detectors, researchers have developed tools that augment teacher abilities to identify and attend to students in need, ultimately improving learning outcomes~\cite{holstein2018student}.

Disengagement represents a well-established area of research at L@S; however, there is an important nuance in our current characterization of disengagement. Existing measures capture what might be more accurately described as \textit{unproductive engagement} rather than avoiding engagement altogether. For instance, mind-wandering involves a shift in attention from task-related processing to task-unrelated thoughts while learners are still engaged in practice~\cite{d2017zone}. In contrast, some forms of disengagement involve avoiding tasks altogether, e.g., delaying to start work, idling, stopping early, or attrition. Throughout this paper, we refer to this pattern as absolute disengagement.

Historically, research has focused on the detection and mitigation of unproductive engagement, with absolute disengagement remaining relatively underexplored\textemdash as there is little to no log trace to help quantify instances of absolute disengagement. However, the growing use of DLPs during structured classwork sessions (e.g., individual seatwork) and recent innovation in session-level measures (see~\cite{gurungdelayedstart2025}) present opportunities to reference peer log trace to detect instances of absolute disengagement. For example, once the majority of students have started working, students who have not yet begun can be considered delayed starters.

We characterize the loss in practice time from instances of absolute disengagement as \textit{coasted time}, and the associated behavior as \textit{coasting behavior}. Coasting behavior comprises 3 sub-behaviors: (a) \textit{delayed start} time lost from the student delaying to start working, (b) \textit{idle} time lost from the student idling mid-practice, and (c) \textit{early stop} time lost from the student stopping early. Understanding coasting is particularly important as DLPs become increasingly integrated into classroom instruction at scale. While platforms offer adaptive learning support and detailed behavioral tracking, their effectiveness is contingent on the effective use of these platforms. If substantial classwork time is lost to coasting, these adaptive systems cannot deliver their intended benefits.

Given the limited prior work on absolute disengagement, we take an exploratory approach to characterize coasting, examine how it varies across student characteristics and school context, and assess its relationship to learning outcomes. This study makes three contributions. First, we operationalize coasting as behavioral measures and extend the current state of the art in disengagement research by incorporating session-level measures~\cite{gurungdelayedstart2025}. Second, we examine variance in coasting behavior across student characteristics and school context, including gender, ethnicity, socio-economic status (SES), English language learner (ELL) status, individualized education program (IEP) status, and locale (urban versus rural). Finally, we investigate the instances of \textit{extra effort}, where students continue practice beyond first assignment completion. We examine whether extra effort predicts better learning outcomes as measured by standardized test performance.

\section{Background}

\subsection{Learning and Time}
Time is a fundamental prerequisite for learning~\cite{fisher2015teaching}. Learning opportunities manifest through a temporal sequence: allocated time enables practice, practice enables problem completion, and accumulated successful problem completions help develop mastery~\cite{koedinger2012knowledge}. Consequently, time is a core component of evaluating the efficient use of learning opportunities~\cite{godwin2021elusive,kovanovic2015does}.

At-scale implementations of digital learning platforms have demonstrated the importance of time-based efficiency metrics. Anderson et al.~\cite{anderson1995cognitive} (see Figure 5), in their summary of a decade-long effort to develop Cognitive Tutor, used time per exercise to demonstrate the efficacy of different pedagogical strategies—including immediate feedback, error flagging, on-demand feedback, and no-feedback conditions. Their work demonstrates the importance of time-normalized measures of progress in providing critical insights into effective instructional design.

Similarly, Koedinger et al.'s~\cite{koedinger2023astonishing} evaluation of 27 datasets reports a promising regularity in learning rates when students productively utilize practice opportunities. Their evaluation spans elementary to college-level students using DLPs for mathematics, science, and language, suggesting that baseline performance differences among learners can be overcome with sufficient access to practice opportunities. These findings underscore a key implication: if practice opportunities drive learning, and available practice time directly influences access to those opportunities, then ensuring the productive use of allocated practice time is essential for facilitating learning.

\subsection{Self-Regulated Learning and Time Use}
Self-regulated learning frameworks conceptualize learning as an active process in which students manage their cognition, motivation, and behavior to achieve learning goals~\cite{zimmerman2002becoming,pintrich2004conceptual,WinneHadwin1998}. Central to effective self-regulation is the strategic management of time and effort~\cite{winne2011cognitive,azevedo2013metacognition}. The effective implementation of these regulatory processes is often observed (and measured) through time, e.g., timely task initiation, sustained engagement despite challenges, and persistence until completion.

Effective self-regulation manifests behaviorally through observable patterns of time use. For example, when faced with difficult content, students with strong self-efficacy persist by allocating more time and effort to overcome the challenge \cite{wolters2003regulation,eccles2020expectancy}. Conversely, difficulties in self-regulation often manifest as temporal dysregulation: procrastination (delaying intended tasks despite negative consequences)~\cite{steel2007nature} and premature task abandonment both represent failures to align time use with learning goals.

Beyond procrastination, time-based behavioral measures that exhibit consistency across extended periods provide valuable insights into learner self-regulatory processes. For instance, research on diligence, characterized by sustained effort and perseverance in learning~\cite{bernard1993diligence,dang2017detecting}, examines patterns of time on task and work completion rates that remain stable over time and predict academic success~\cite{galla2014academic}. The temporal stability of these measures underscores the value of consistent behavioral indicators in structured practice settings, where repeated observations over longer time periods can reveal how students manage available learning opportunities.

\subsection{Disengagement in Digital Learning Environments}
Insights from SRL research and studies on productive use of learning time have informed the development of detectors for disengagement in DLPs. These detectors incorporate raw or aggregated temporal measures to distinguish productive from unproductive engagement. Measures for gaming the system~\cite{baker2008developing,paquette2014towards}, wheel-spinning (unproductive persistence)~\cite{beck2013wheel,gong2015towards}, off-task behavior~\cite{baker2004off}, and mind-wandering~\cite{christoff2018mind,mills2016automatic} use time as a critical component to capture various forms of disengagement. For instance, rapid help-seeking within seconds of encountering new content can signal attempts to game the system rather than a genuine effort to learn~\cite{paquette2014towards}. Similarly, response-time patterns have been used to infer low effort in help use~\cite{shih2011response,gurung2021examining}.

These within-task disengagement measures have demonstrated meaningful relationships with learning outcomes. Prior work has reported that students exhibiting low effort on DLP-provided help are significantly more likely to engage in wheel-spinning~\cite{gurung2021examining}. Similarly, students who go on to pursue non-STEM fields exhibit substantially higher rates of gaming the system behavior than their STEM-pursuing peers~\cite {san2014predicting}. These findings underscore how temporal patterns during active platform use reveal important aspects of self-regulation and engagement.

Most existing measures capture disengagement during active system use. While within-task disengagement captures important aspects of opportunity loss, it overlooks opportunity loss from students who avoid engagement altogether. This gap is particularly salient in structured classwork sessions, where teachers allocate predefined intervals of practice time with clear expectations for student engagement~\cite{asri2017academic,brookfield2015skillful}. In such settings, session-level measures~\cite{gurung2021examining} of when students start work, how long they remain engaged, and when they choose to stop reflect self-regulatory processes including effort regulation~\cite{wolters2003regulation}, goal pursuit~\cite{locke2019development}, and response to contextual factors~\cite{pintrich2004conceptual,WinneHadwin1998}. Examining these temporal patterns relative to peers provides complementary insights into how students manage allocated learning opportunities within the classroom context.

\subsection{Demographic Patterns in Engagement}
Societal limitations and extrinsic factors can impact the productive utilization of available opportunities. Baker et al.\cite{baker2025gaming} report gender differences in gaming the system behavior in a digital learning game, with female students engaging in such behavior much less frequently than their male counterparts. This difference in gaming the system mediated the relationship between gender and learning outcomes, potentially attributable to differences in decision-making, interpersonal goals versus task orientation~\cite {koivisto2014demographic}, or gender-based differences in motivation and cognitive strategies~\cite{wolters1998contextual}. Similarly, racial and ethnic differences in time use have been documented in homework completion and academic engagement~\cite{dunatchik2022racial}, with family socioeconomic background accounting for differences in homework time between Hispanic and White students and partially explaining differences between Black and White students. Research on time as a resource reveals that Black and Hispanic students experience significantly greater time poverty than White or Asian students~\cite{wladis2024,wladis2024s}, reflecting competing demands that constrain time available for academic pursuits.

The benefits of learning opportunities can also vary across learners due to qualitative differences in learner needs and readiness to leverage them effectively. Students from lower socioeconomic backgrounds face resource constraints and competing time demands that affect their capacity to dedicate time to academic work. Among students receiving additional services (IEP) or language support (ELL), a lack of support can impede IEP learners' ability to effectively use DLPs~\cite{kim2022increasing,thomas2024neglected}, and a lack of English language proficiency can impact learner performance and motivation~\cite{rentas2025exploring}. Geographic context also shapes engagement opportunities, with rural students facing resource constraints, limited access to specialized support, and persistent achievement gaps compared to urban peers~\cite{zhang2015education}. Despite documented achievement gaps across these demographic groups, less is known about how these factors shape temporal patterns of engagement during structured practice opportunities.

% \subsection{Digital Platforms and Classwork Implementation}

% Growing adoption of digital learning platforms during classwork has transformed student engagement with practice. Structured sessions during regular class periods increasingly rely on platforms for individualized practice. Depending on context, sessions can be implemented as instruction followed by practice, practice entirely, or practice combined with reflection. This shift introduces opportunities and challenges: platforms provide immediate feedback and adaptive sequencing, but reduced teacher oversight may create disengagement opportunities.

% Classwork sessions are typically predetermined and bounded, with some level of teacher control over timing and structure of practice intervals. 
% % For instance, some teachers might dedicate an entire period to classwork practice on a DLP, whereas others might prefer to divide the session into parts with teacher instruction followed by classwork activities, or classwork activities followed by review. 
% Recent work by~\citeN{gurungdelayedstart2025} leveraged this structure to propose an algorithm that mines logs across students to dynamically infer practice intervals during classwork sessions, enabling detection of absolute disengagement at the session level (i.e., avoiding work by delaying the start and stopping early).  

\section{Defining Coasting Behavior}
Digital platforms offer clear pathways for translating practice time into learning progress: productive use of practice time affords more problem attempts, supporting lesson completion and, in some systems, mastery. Productive practice requires, at a minimum, that students use the allocated time. However, without productive use, the allocated time does not translate into learning progress.

\begin{figure*}[!ht]
  \Description{A problem-level visualization of student engagement during a classwork session (taken and modified from Gurung et al.~\cite{gurungdelayedstart2025}). The (a) class session time represents the overall duration of the classwork session. (b) Student session time shows an individual student's active practice period. The figure also illustrates (c) absence, (d) delayed start, (e) early stop, and (f) idling behavior.}
  \centering
  \includegraphics[width=0.9\linewidth]{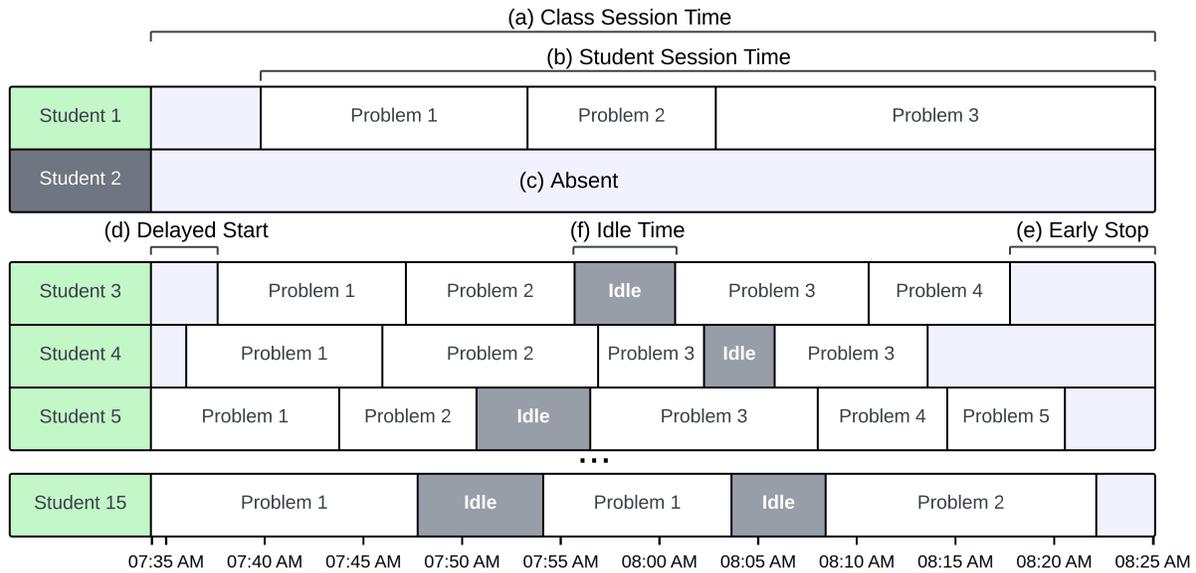}
  \caption{A problem-level visualization of student engagement during a classwork session (inspired by visualization from Gurung et al.~\cite{gurungdelayedstart2025}). The (a)~class session time represents the overall duration of the classwork session. (b) Individual student's active practice time. The figure also illustrates (c) absence, (d) delayed start, and (e) early stop, and (f) idle time.}
  \label{fig:class-session}
\end{figure*}

As illustrated in Figure~\ref{fig:class-session}, even during structured classwork sessions when all students are expected to work, practice time can be lost from students avoiding work. This includes delays in starting work (Figure~\ref{fig:class-session} d), stopping work early (Figure~\ref{fig:class-session} e), and idling mid-practice (Figure~\ref{fig:class-session} f)~\cite{baker2004off}. Delayed start captures time lost at the beginning of a session. Idle time captures periods of inactivity mid-practice. Early stop captures time lost when students stop work before the session ends, despite available practice time. As such, we define this lost practice time as \textit{coasted time} and the associated behavior as \textit{coasting behavior}. Formally,

\begin{equation}
\text{Coasted Time} = \text{Delayed Start} + \text{Idle Time} + \text{Early Stop}
\end{equation}

We use coasting as an umbrella term because delaying starts, idling mid-session, and early stopping all reflect the same underlying phenomenon: underutilization of available practice time. This definition intentionally distinguishes coasting from absence, as classwork practice is not available to students who are absent. This terminology highlights a time-based definition of opportunity loss that is agnostic to the specific cause (e.g., distraction, avoidance, or competing demands) and enables consistent measurement of these behaviors using platform log data. 
\section{Method}

\subsection{Dataset}

This study analyzes a publicly available\footnote{\scriptsize{\href{https://www.openicpsr.org/openicpsr/project/238826/version/V2/view}{https://www.openicpsr.org/openicpsr/project/238826/version/V2/view}}} dataset containing approximately 3 million transaction logs from ASSISTments~\cite{feng2025empowering}, a DLP that provides curriculum-aligned mathematics content with immediate feedback and on-demand help~\cite{heffernan2014assistments}. The dataset spans two school years (2022--2023 and 2023--2024) and includes data from 1,431 students across 71 classes with 34 teachers in 22 schools across 17 U.S. states. Schools that did not regularly use ASSISTments during classwork sessions were excluded from the analysis. The average class size was 20 students (SD=4.86). In addition to log data, the dataset includes scores from a Measure of Academic Progress (MAP~\cite{he2021map}) assessment administered in the fall (pre-test) and spring (post-test) of each school year.

The dataset includes student demographic information. In percentages, 58.63\% of students in the sample were White, 22.01\% Hispanic, 8.50\% African American, and 6.57\% were categorized as Other (American Indian, Asian, or Native Hawaiian); 4.33\% of race/ethnicity data were missing. By gender, 45.49\% of students were male and 49.62\% were female, with 4.89\% missing. For socio-economic status, 30.19\% of students were eligible for free or reduced-price lunch (FRL; used as a proxy for SES) and 41.10\% were not eligible; FRL status was missing for 28.72\% of students. Regarding special education, 9.43\% of students had an IEP and 72.68\% did not, with 17.89\% missing. For English learner status, 8.39\% of students were classified as ELL and 66.95\% were not, with 24.67\% missing. Schools were located in urban (35\%) and rural (65\%) areas. Although missingness is substantial for some administrative indicators (notably FRL, IEP, and ELL), this is not uncommon in multi-site datasets collected across 17 states, where reporting practices and data availability can differ.

Data collection followed IRB-approved protocols, and participation was conducted under an opt-out informed consent process.

\subsection{Session Identification}

\begin{table*}[!hb]
\centering
\caption{Descriptive Statistics by Session Type}
\label{tab:session_types}
\begin{tabular}{lcccc}
\hline
Session Type & \begin{tabular}{@{}c@{}}Total\\Student-Sessions\end{tabular} & \begin{tabular}{@{}c@{}}Inferred\\Class-Sessions\end{tabular} & \begin{tabular}{@{}c@{}}Avg Students\\per Session\end{tabular} & \begin{tabular}{@{}c@{}}Avg Session\\Length (mins)\end{tabular} \\
\hline
Independent Work & 20,415 & 15,011 & 1.36 (0.75) & 7.23 (9.07) \\
Classwork & 47,188 & 3,658 & 12.92 (5.79) & 25.88 (16.90) \\
Homework & 7,831 & 6,581 & 1.19 (0.83) & 10.80 (14.62) \\
\hline
\end{tabular}
\end{table*}

We applied the session inference algorithm proposed by Gurung et al.~\cite{gurungdelayedstart2025} (see Algorithm 1) to classify student practice into three categories: classwork sessions, at-school independent work, and homework sessions. The algorithm uses temporal patterns in student log data and active-student counts to distinguish session types. As the schools are based in the United States, we use the local timezone to define school hours as 7 a.m. to 3 p.m. Sessions during school hours with more than five concurrently active students are classified as classwork sessions, whereas sessions during school hours with fewer than five active students are classified as at-school independent work. For homework, practice sessions outside school hours or on weekends are classified as homework.

Table~\ref{tab:session_types} presents descriptive statistics of the inferred sessions. Most of the student practice occurred during classwork sessions (N=47,188 student-sessions), with an average classwork session length of 25.88 minutes ($SD=16.90$) and a session size of 12.92 ($SD=5.79$) students per session.

\subsection{Measures}

\subsubsection{Session-level Time Based Measures}

We use session-level measures to contrast individual student sessions with the overall classwork session. As illustrated in Figure~\ref{fig:class-session}, key measures include the (a) classwork session length, (b) individual student session length, (d) delayed start, and (e) early stop.

\begin{table}[!h]
\centering
\caption{Time-based measures used to explore the opportunity loss from coasted time.}
\label{tab:gender_ethnicity}
\begin{tabular}{p{0.23\linewidth} p{0.68\linewidth}}
\hline
Measure & Description \\
\hline
Delayed Start & Time lost from the student delaying the start of work relative to the classwork session start time. \\
Idling & Time lost from inactivity during practice. Following Gurung et al.~\cite{gurungdelayedstart2025}, periods of inactivity exceeding the 99th percentile of response times (1.8 minutes, rounded to 2 minutes) are classified as idle time.\\
Early Stop & Time lost from the student stopping work before the end of the classwork session. \\
Total Coasted Time & The sum of delayed start, idle time, and early stop, representing the practice time lost to coasting behavior during a classwork session. \\
\hline
\end{tabular}
\end{table}

We aggregate these measures across classwork sessions for each student to compute individual-level means.

\subsubsection{Extra Effort}
ASSISTments~\cite{heffernan2014assistments} is a teacher-paced system in which teachers control assignments and typically sequence work according to their preferred curricular order. Depending on instructional goals and logistical flexibility, teachers may release additional content for students who finish early, or allow students to use remaining classwork time to revisit previously incomplete work.

We define \textit{extra effort} as continued practice beyond the first assignment completion during classwork sessions. Because ASSISTments logs record assignment completion, we identify extra effort as instances where the student continued working on additional assignments after completing the first required assignment during a classwork session. In this context, starting additional work reflects engagement beyond minimal compliance with the initially assigned task.

\subsection{Analysis}
\subsubsection{Coasted Time}
We evaluate lost practice opportunity by contrasting available classwork session time with the time the student spent practicing mathematics. We then examine the time lost from coasting due to the student delaying to start, idling mid-practice, and stopping early.

Depending on teachers’ pedagogical needs and instructional choices, the use of ASSISTments for supplemental practice can vary in frequency and approach across classes. To account for the variance and enable comparisons across classes, we evaluate coasted time as a proportion of available classwork time per session. 

Not all early stopping necessarily reflects disengagement, as students may stop after completing their assigned work rather than avoiding practice. To account for this, we compute an adjusted coasted time by setting the early stop to zero for sessions in which students completed at least one assignment. This yields a conservative estimate of coasting that attributes early stopping in these sessions to task completion rather than avoidance of available practice time.

% Additionally, as stopping after completing the first assignment may be acceptable in some classroom contexts. We therefore evaluate early stopping in two ways: (1) as lost time from stopping before the end of the classwork session, regardless of completion status, and (2) as zero if the student completed at least one assignment before stopping.

\subsubsection{Temporal stability of time-based measures.} We evaluate the temporal stability (reliability) of time-based measures using a Generalizability Theory (G-Theory~\cite{cronbach1972dependability,brennan1992generalizability,brennan2010generalizability}). Specifically, we assess the consistency of students’ delayed start, idle time, and early stopping across classwork sessions to examine their reliability over time. 

We specified the variance component model for each session-level measure using Equation~\ref{eqn:gtheory_eqn}. The model uses random intercept to account for individual differences $\gamma_j^{(student)}$, variability across months $\gamma_l^{(month)}$ and varaince in individual difference across months $\gamma_{jl}^{(student:month)}$ for each measure $Y_{jl}$. 

\begin{equation}
\begin{aligned}
\label{eqn:gtheory_eqn}
 Y_{jl} &= \beta_0 + \gamma_j^{(student)} + \gamma_{l}^{(month)} +  \gamma_{jl}^{(student:month)} + \epsilon_{}
\end{aligned}    
\end{equation}

We computed G-coefficients (Equation~\ref{eqn:g-coefficient}) to assess the overall reliability, also called generalizability, of each measure and $\phi$-coefficients (Equation~\ref{eqn:phi-coefficient}) to determine their consistency for decision-making applications.

\begin{equation}
\label{eqn:g-coefficient}
G = \frac{\sigma^2_{student}}{\sigma^2_{student} + \sigma^2_{relative}}    
\end{equation}

\begin{equation}
\label{eqn:phi-coefficient}
\phi = \frac{\sigma^2_{student}}{\sigma^2_{student} + \sigma^2_{absolute}}
\end{equation}

We follow Brennan's~\cite{brennan2010generalizability} framework, which associates G-coefficients above 0.80 with strong reliability suitable for high-stakes individual decisions. Values between 0.60 and 0.80 indicate moderate reliability that is appropriate for research purposes and group-level comparisons. Values below 0.60 suggest lower reliability, indicating substantial context-dependence of the measure.

\subsubsection{Student Characteristics and School Context}
We examine the variance in student coasting across student characteristics and school contexts using multilevel linear regression, while controlling for prior performance and accounting for class-level variance as outlined in Equation~\ref{eqn:variance-in-coasting1}. Due to potential feature collinearity and varying patterns of missing data across demographic variables, we estimated separate models for: (1) gender and ethnicity with their interaction; (2) SES; (3) IEP status; (4) ELL status; and (5) school urbanicity. 

\begin{equation}
\begin{aligned}
\label{eqn:variance-in-coasting1}
\text{Y}_{i} &= \beta_{0} + \beta_{1} {\text{ prior performance}_{j}} + \\ 
&\quad \sum_{k=1}^{n-1} \beta_{2k} {\text{ student characteristic}_{k}} + \\
&\quad \gamma_{n}^{class} + \varepsilon_{},   
\end{aligned}
\end{equation}

\subsubsection{Extra Effort}
We examine the benefit of extra effort on student performance on standardized tests using a multilevel regression, while controlling for prior standardized test performance, time on task, and accounting for class-level variance as outlined in Equation~\ref{eqn:variance-in-coasting}. Because classes varied in the frequency of classwork sessions, both time on task and extra effort were measured as proportions of available classwork time to enable standardized comparisons across classes.

\begin{equation}
\begin{aligned}
\label{eqn:variance-in-coasting}
\text{Y}_{i} &= \beta_{0} + \beta_{1} {\text{ prior performance}_{j}} + \beta_{2} {\text{ time on task}_{j}} \\ 
&\quad \beta_{3} {\text{ extra effort time}_{j}} + \gamma_{n}^{class} + \varepsilon_{},   
\end{aligned}
\end{equation}

We evaluate improvements in model fit from extra effort using the Bayesian Information Criterion (BIC) and R-squared measures. We follow Raftery's~\cite{raftery1995bayesian} guidelines for BIC to evaluate the model performance: a BIC difference of $0\text{-}2$ indicates `weak'; $2\text{-}6$ `positive'; $6\text{-}10$ `strong'; and $>10$ `very strong' evidence in favor of the model with the smaller BIC. 

\section{Results}

\subsection{Extent and Composition of Coasting Behavior}
As reported in Table~\ref{tab:reliability_analysis}, our evaluation of the temporal stability of the session-level measures indicates that delayed start and early stopping exhibit moderate reliability (G-coefficients = 0.73 and 0.71, respectively; $\phi$ = 0.72 and 0.70). In contrast, idle time shows lower reliability (G = 0.53; $\phi$ = 0.53), suggesting that idling may be more context-dependent and sensitive to features of the assigned content (e.g., problem difficulty or task structure) in a given session.

\begin{table}[!h]
\caption{Evaluating the temporal stability (reliability) of student behavior that contributes to coasting behavior.}
\label{tab:reliability_analysis}
\begin{tabular}{lll}
\hline
                               & \multicolumn{2}{c}{ASSISTments}                 \\
\multicolumn{1}{c}{Measure}    & \multicolumn{1}{c}{$G$}        & \multicolumn{1}{c}{$\phi$}        \\
\hline                               
delayed start (mins)           &0.73                          &0.72                          \\
idle time (\textgreater 2mins) & 0.53                         & 0.53                         \\
early stop (mins)              &0.71                          &0.70                          \\
\hline
\end{tabular}
\end{table}

Our evaluation of coasting time is reported in Table~\ref{tab:coasting_descriptives}. On average, classwork sessions lasted 26.9 minutes ($SD = 7.93$). Of the available classwork session time, students actively practiced mathematics in ASSISTments for an average of 10.8 minutes ($SD = 4.26$), representing 40.15\% of available time. The remaining 59.85\% (16.06 minutes) was lost to coasting.

\begin{table}[!h]
\centering
\caption{Descriptive Statistics for Coasting Behaviors}
\label{tab:coasting_descriptives}
\resizebox{1\linewidth}{!}{%
\begin{tabular}{lccccc}
\hline
Measure (mins) & Mean & SD & Median & \% Available & \% Coasted \\
\hline
Available Time & 26.9 & 7.93 & 26.1 & 100\% & — \\
Time on Task & {10.8} & {4.26} & {9.94} & {40.15\%} & — \\
\hline
Total Coasted Time & {16.06} & — & — & {59.85\%} & {100\%} \\
\quad Delayed Start & 5.75 & 3.15 & 5.00 & 21.38\% & 35.79\% \\
\quad Idle Time (>2)    & 0.31 & 0.45 & 0.18 & 1.12\% & 1.96\% \\
% \quad Idle Time\\(>2 mins) & 0.314 & 0.45 & 0.18 & 1.12\% & 1.96\% \\
\quad Early Stop & 10.0 & 4.48 & 9.41 & 36.18\% & 62.25\% \\
\hline
\multicolumn{6}{l}{\scriptsize{\% Available = percentage of total available time; \% Coasted = percentage of total coasted time.}}
\end{tabular}}
\end{table}

Early stopping accounted for 62.25\% of coasted time, followed by delayed start (35.79\%), with idle time contributing minimally (1.96\%). Results suggest that most lost practice time occurs at the session boundaries rather than during practice itself. Even after setting the early stop to zero for students who completed at least one assignment, students still coasted for 32\% of the available time.

\subsection{Coasting Across Student Characteristics and School Contexts}
Given that coasting represents a meaningful loss of practice opportunity, we next examine whether student-level characteristics and school context are associated with differences in coasting behavior. As we are evaluating the proportion of coasted time relative to the available classwork time, lower values are better. 

\subsubsection{Gender and Race/Ethnicity}
Table~\ref{tab:gender_ethnicity} reports a two-level linear mixed-effects model predicting total coasted time from Fall MAP score, gender, race/ethnicity, and their interactions, with students nested within classes and a random intercept for class. 

\begin{table}[!h]
\centering
\caption{Examining the difference in student coasted time across gender and race/ethnicity.}
\label{tab:gender_ethnicity}
% \resizebox{1\linewidth}{!}{%
\begin{tabular}{lcc}
\hline
 & \begin{tabular}{@{}c@{}}Model 1 \\ (Unadjusted)\end{tabular}& \begin{tabular}{@{}c@{}}Model 2 \\ (Adjusted)\end{tabular} \\
Predictor & $\beta$ & $\beta$ \\
\hline
Intercept & 0.59 *** & 0.32 *** \\
MAP Score (Fall) & 0.01 ** & -0.04 *** \\
Gender (Male) & 0.02 *** & 0.00 \\
Ethnicity: African American & 0.00 & 0.02 \\
Ethnicity: Hispanic & -0.01 & 0.01 \\
Ethnicity: Other & -0.01 & 0.02 \\
African American $\times$ Male & 0.01 & -0.02 \\
Hispanic $\times$ Male & 0.00 & 0.00 \\
Other $\times$ Male & -0.05 ** & -0.03 \\
\hline
\multicolumn{3}{l}{\textit{Random Effects}} \\
$\sigma^2$ & 0.01 & 0.01 \\
$\tau_{00}$ & 0.00 & 0.01 \\
ICC & 0.36 & 0.41 \\
\hline
$N$ (Classes) & 68 & 68 \\
$N$ (Students) & 1353 & 1353 \\
Marginal $R^2$ / Conditional $R^2$ & 0.030 / 0.380 & 0.112 / 0.476 \\
\hline
\multicolumn{3}{l}{\scriptsize $*p < .05$, $**p < .01$, $***p < .001$.}
\end{tabular}%}
\end{table}

For unadjusted coasting behavior (Model 1), higher prior performance was associated with slightly greater coasted time ($1\%$, $p < .01$). A 1\% change in coasting time corresponds to 32.28 seconds. Male students on average had higher coasted time than female students ($2\%$, $p < .001$). We did not observe a significant difference in main effects between ethnicities (African American, Hispanic, Other; relative to the reference group, White students). Interaction terms were also largely non-significant, except for the interaction between male students and the other group, which was negative ($-5\%$, $p < .01$), indicating a smaller male–female difference in coasted time for students categorized as “Other” compared to the reference group (White Female). The intraclass correlation (ICC = 0.36) indicates that a 36\% of the variance in coasted time across classes. Fixed effects explained a modest share of variance (marginal $R^2 = 0.03$), while the full model, including class-level effects, accounted for $R^2 = 0.38$ (conditional).

When we adjust for assignment completion (Model 2; early stop = 0 for $\geq1$ assignment completion), previously observed differences by gender and race/ethnicity are no longer statistically significant. This attenuation suggests that between-group differences in unadjusted coasted time are largely attributable to differences in student pace rather than disparities in engagement. 
% Put differently, the group differences in unadjusted coasting reflect differential completion patterns more than differences in loss of opportunity due to coasting.

\subsubsection{ELL Status, SES, IEP Status and Urbanicity} Similarly, we examined whether coasting behavior differed by ELL Status, SES, IEP Status and urbanicity (controlling for prior achievement and class level variance). Students with an IEP exhibited slightly lower coasting than their non-IEP peers (-2\%, p<.05); no statistically significant differences were observed for SES, ELL status, or urbanicity. When adjusting for assignment completion (early stop = 0 for $\geq$ 1 assignment completion), the IEP–non-IEP difference in coasting behavior was no longer statistically significant. The comparison across groups is illustrated in Figure~\ref{fig:comparisions}.

\begin{figure*}[!ht]
  \Description{Average proportion of coasted time to available classwork session time by student characteristics and locale. The top row contains unadjusted coasted time; the bottom row is adjusted for assignment completion (early stop = 0 for $\geq$1 assignment completion). Error bars represent 95\% CIs.}
  \centering
  \includegraphics[width=.8\linewidth]{images/coasted_plot.jpeg}
  \caption{Average proportion of coasted time to available classwork session time by student characteristics and locale. We only observed a significant difference in coasted time between IEP and non-IEP students. The top panel contains unadjusted coasted time; the bottom panel is adjusted for assignment completion (early stop = 0 for $\geq$1 assignment completion). Error bars represent 95\% CIs.}
  \label{fig:comparisions}
\end{figure*}

\subsection{Extra Effort and Achievement}
Next, we evaluated whether extra effort predicted Spring MAP performance, controlling for time on task, prior performance (Fall MAP), and class as a random intercept to account for between-class variance. Spring MAP scores were standardized, and both time on task and extra effort were measured as proportions of available classwork time (range 0–1).

Table~\ref{tab:achievement_models} reports the multilevel models with class specified as a random intercept. In Model 3, Fall MAP score strongly predicted Spring MAP ($0.78$, $p < .001$), whereas time on task was not a statistically significant predictor ($-0.02$, $p = .919$). 

\begin{table}[!h]
\centering
\caption{Testing whether extra effort predicts standardized test performance beyond time on task and prior performance, accounting for between-class variance.}
\label{tab:achievement_models}
\begin{tabular}{lcc}
\hline
 & \multicolumn{1}{c}{Model 3} & \multicolumn{1}{c}{Model 4} \\
Predictor & $\beta$ & $\beta$ \\
\hline
Intercept & -0.01 & -0.01 \\
MAP Score (Fall) & 0.78 *** & 0.77 ***\\
Time on Task Proportion & - 0.02 & -0.20 \\
Extra Effort Proportion & & 1.44 ***\\
\hline
\multicolumn{3}{l}{\textit{Random Effects}} \\
$\sigma^2$ & 0.23 & 0.22\\
$\tau_{00}$ & 0.06 & 0.05\\
ICC & 0.21 & 0.20\\
\hline
$N$ (Classes) & 71 & 71 \\
$N$ (Students) & 1425 & 1425 \\
Marginal $R^2$ / Conditional $R^2$ & 0.683 / 0.749 & 0.695 / 0.755 \\
$\Delta R^2$ (Marginal) & — & 0.012 \\
$BIC$  & 2098 & 2087 \\
\hline
\end{tabular}
\end{table}

The inclusion of extra effort (Model 4) improved model fit, as indicated by a reduction in BIC (from 2098 to 2087) and an increase in marginal $R^2$ from 0.683 to 0.695 ($\Delta R^2 = 0.012$). Following Raftery's guidelines~\cite{raftery1995bayesian}, an 11-point reduction in BIC is considered very strong evidence of improved model performance from the addition of extra effort measure (Model 4). Extra effort was a significant positive predictor of Spring MAP ($\beta = 1.44$, $p < .001$). As extra effort is a proportional measure, a 0.10 (10\%) increase in extra effort corresponds to an expected 0.14 SD increase in Spring MAP, holding other covariates constant. The ICCs (0.20--0.21) indicate that roughly one-fifth of the variance in Spring MAP scores was attributable to between-class differences.

\section{Discussion}

\subsection{Opportunity Loss During Classwork}

We find that only 40\% of available classwork time was dedicated to active practice. This substantial gap between intended and actual learning opportunity highlights how pervasive opportunity loss is during structured classwork sessions. Early stopping accounted for 62\% of coasted time, delayed starts for 36\%, and idle time for just 2\%. The low proportion of idle time suggests that once students begin working, they tend to remain engaged, i.e., the loss of practice time primarily occurs at session boundaries, not during practice itself. Generalizability analyses~\cite{cronbach1972dependability,brennan2010generalizability} support this interpretation: delayed start and early stop showed moderate reliability across sessions (G=0.73 and 0.71, respectively), suggesting these are consistent behavioral patterns, whereas idling was less reliable (G=0.53), indicating greater sensitivity to session or content characteristics.

To contextualize the reliability and predictive validity of session-level measures more broadly, psychometric measures such as the Big Five personality traits typically achieve \textit{G}-coefficients between 0.80 and 0.90 and \textit{$\phi$}-coefficients between 0.75 and 0.85~\cite{arterberry2014application}. Our delayed start (G = 0.73, $\phi$ = 0.72) and early stop (G = 0.71, $\phi$ = 0.70) measures approach these benchmarks, demonstrating strong reliability for behavioral measures in dynamic learning contexts.

Critically, coasted time remains substantial (32\%) even after adjusting for assignment completion, indicating that opportunity loss is not simply a consequence of efficient students finishing early. Students leave available practice time unused even when additional work is available. Coasted time captures time not spent on math practice on the learning platform and does not necessarily indicate unproductive time. Students could be utilizing coasted time towards learning activities beyond the DLP, such as receiving teacher instruction or working with peers. However, the positive association between extra effort and learning outcomes suggests that at least some of this time represents missed opportunities for productive practice. This disconnect between available opportunity and actual engagement underscores the need for improved practitioner involvement and platform affordances~\cite{an2020ta, holstein2018student, martinez2012orchestrating, martinez2013mtclassroom, fisher2015teaching} that support clearer expectations and better monitoring of student engagement during classwork\textemdash a clear opportunity for our research community to develop and evaluate.

\subsection{Demographic Patterns in Coasting}

We observed significant differences in coasting by gender and IEP status, but not by race/ethnicity, SES, ELL status, or school locale. Male students coasted significantly more than female students, by approximately 32 seconds per session. While modest per-session, this difference compounds over repeated classwork sessions throughout the school year. These findings are consistent with those of Baker et al.~\cite{baker2025gaming}, who found that male students exhibit higher levels of disengagement (i.e., gaming the system) than the female students. However, in our analysis, this difference was no longer significant after adjusting for assignment completion, suggesting that the gender gap reflects differences in the pace of assignment completion rather than differences in disengagement. Male students, on average, completed assignments faster, resulting in more recorded early stop times before adjustment. Once assignment completion is accounted for, male and female students show comparable engagement patterns. The significant difference in unadjusted coasting time appears to be driven by male students completing assignments faster on average, which inflates their recorded early stop time.

Unlike prior findings that students with IEPs struggle to productively engage with DLPs to derive benefits~\cite{kim2022increasing,thomas2024neglected}, we find that students with IEPs coasted significantly less than their non-IEP peers. This counterintuitive finding may reflect several factors, including greater teacher monitoring and support (e.g., more frequent check-ins), which could reduce opportunities for disengagement. In addition, IEP accommodations may involve modified or more manageable workloads, and/or students may progress at a slower pace, which would naturally reduce early stopping and thus lower total coasted time.

The absence of differences by race/ethnicity, SES, ELL status, and location is noteworthy. Coasting appears relatively universal across these groups, indicating that broadly deployed interventions may be effective in reducing coasting; nonetheless, implementation should remain sensitive to local context, including teachers’ classroom practices and instructional decisions~\cite{asri2017academic,brookfield2015skillful} (e.g., how practice time is structured, monitored, and reinforced).

\subsection{Implications for Research}
Extra effort during classwork sessions was a significant predictor of post-test performance above and beyond overall time on task. This pattern suggests that extra effort reflects more than simply spending additional time or increasing content exposure during classwork. Continued practice beyond assignment completion likely reflects motivational and self-regulatory characteristics that directly facilitate learning~\cite{zimmerman2002becoming,pintrich2004conceptual}, such as stronger mastery orientations~\cite{wolters2003regulation, locke2019development, galla2014academic} or greater intrinsic motivation~\cite{gollwitzer2006implementation}. The null effect of time spent working toward initial completion supports this interpretation: if time on task directly translated to learning through exposure, all practice time would predict performance on standardized tests. The fact that only extra effort predicted post-test performance points to qualitative differences in engagement beyond differences in students' use of available practice time, as captured by time on task.

This study extends disengagement research by introducing a complementary focus on task avoidance. While prior work emphasized within-task behaviors (gaming the system~\cite{baker2008developing,paquette2014towards,baker2008students}, wheel-spinning~\cite{beck2013wheel,gong2015towards}, mind-wandering~\cite{mills2016automatic,christoff2018mind}), coasting builds on prior work on task avoidance (e.g., off-task behavior~\cite{baker2004off}, procrastination~\cite{steel2007nature,borchers2024you}) by incorporating session-level measures to identify loss of practice time from avoiding engagement.

\subsection{Implications for Practice}
As DLPs are increasingly adopted at scale, the current findings underscore the need for teachers to set clear and explicit expectations for sustained practice during classwork sessions. Students may benefit from clear expectations to continue practicing throughout allocated time, supported through active monitoring, extension activities for early finishers, or incentive structures that reward sustained practice~\cite{asri2017academic,brookfield2015skillful}. Platform designers can reinforce these norms at scale by making engagement more visible and actionable~\cite{an2020ta,holstein2018student,martinez2012orchestrating,martinez2013mtclassroom} (e.g., progress indicators, prompts for inactivity, and readily accessible ``what to do next” pathways that encourage continued practice). At the school level, these results also underscore the importance of better implementation norms when evaluating the integration of learning platforms for supplemental practice, as even effective platforms will yield limited returns if there is substantial coasting during routine use.

A key advantage of the current approach to analyzing coasting behavior is the use of easily interpretable \textit{system-general} session-level measures, i.e., delayed start, mid-session idleness, and early stopping. Our implementation of session-level measures in ASSISTments demonstrates their generalizability by extending them beyond MATHia (formerly Cognitive Tutor) and iReady, as implemented by Gurung et al.~\cite{gurungdelayedstart2025}, to a new platform. As session-level measures and idleness require log traces that most systems capture, the measures required for coasting should be easily calculated for most learning platforms.

\section{Limitations and Future Directions}
Our analyses utilize a data-driven strategy~\cite{gurungdelayedstart2025} that uses student trace logs to dynamically approximate session boundaries for establishing class start and end times. While pragmatic and scalable, this approach is sensitive to certain edge cases that can introduce measurement uncertainty. For example, if some students consistently begin early or continue working beyond the end of the classwork session, inferred session bounds may be extended, which could affect delayed-start and early-stop estimates for other students in the same class. Future work linking platform logs to bell schedules or teacher-defined session windows could quantify the accuracy of the inferred boundaries and further refine coasting estimation.

Our current evaluation is correlational, so the observed association between coasting and achievement should be interpreted as such. Although we adjust for prior achievement, unmeasured factors such as self-regulation~\cite{wolters1998contextual}, diligence~\cite{galla2014academic}, procrastination~\cite{steel2007nature}, and goal orientation~\cite{locke2019development} may influence both coasting and learning outcomes. Establishing a causal relationship will require more rigorous analysis (e.g., quasi-experimental designs or randomized studies) to test whether interventions that reduce coasting lead to subsequent gains in achievement. Additionally, coasting is operationalized as a quantitative measure of time and does not capture the qualitative mechanisms of disengagement. Future work should examine the motivational and self-regulatory factors~\cite{winne2011cognitive,azevedo2013metacognition,wolters2003regulation,eccles2020expectancy} that drive coasting behavior, as well as how coasting co-occurs with other forms of disengagement (e.g., gaming the system~\cite{baker2008students,paquette2015cross}, mind-wandering~\cite{christoff2018mind,mills2016automatic}, wheel-spinning~\cite{beck2013wheel,gong2015towards}, effort~\cite{gurung2021examining,shih2011response}), to develop a more comprehensive account of student engagement with digital learning platforms and their impact on learning.

Our sample was drawn from 71 middle school classrooms using ASSISTments~\cite{heffernan2014assistments} for mathematics, and coasting patterns may differ across grade levels, subject domains, instructional settings, and platforms. Replication in more diverse contexts is needed to establish generalizability. Additional validation work is also warranted to strengthen inferences about the mechanisms underlying coasting~\cite{kane2013validating}. For example, qualitative and mixed-methods studies, including surveys, student interviews, and experience-sampling approaches, could clarify the motivations and decision-making processes that drive delayed starts, idling or off-task behavior, and early stopping. Coasting is also likely shaped by class-level contextual factors~\cite{pintrich2004conceptual,winne2011cognitive}. Factors such as teacher monitoring and feedback, peer norms, task design, and incentive structures may moderate the extent of task avoidance and opportunity loss. Future research should explicitly measure these contextual features and examine their role using multilevel designs that model school-level, teacher-level, and classroom-level variation, to clarify when and why coasting varies across settings.

\section{Conclusion}
Effective use of learning opportunities is a fundamental determinant of learning, yet much of the time allocated to classwork can be lost to disengagement. This study introduces coasting behavior as a framework for quantifying absolute disengagement during classroom DLP use, operationalized as the sum of delayed start, idling, and early stop time. Across 1,431 students from 22 schools in 17 U.S. states who used ASSISTments for math practice throughout the academic year, students spent only 40\% of available classwork time on active practice, with early stopping accounting for the majority of lost practice time. Coasting did not vary meaningfully across student demographics or school locale, suggesting that it is a pervasive phenomenon rather than one concentrated in specific populations. Critically, continuing work beyond the first assignment (extra effort) significantly predicted standardized test performance beyond time on task alone, indicating that voluntary continued practice may capture self-regulatory and motivational aspects of engagement.

Theoretically, coasting extends disengagement research by demonstrating that disengagement during practice operates at multiple levels: the session level, reflected in delayed starts and early stops, and the task level, reflected in idle time. This multilevel analysis provides a more comprehensive account of how students underuse the allocated practice time. Building on prior work~\cite{gurungdelayedstart2025}, this study further establishes the cross-system generalizability of session-level measures by extending them from Cognitive Tutor and iReady to ASSISTments. Practically, the prevalence of coasting points to opportunities for platform improvements that strengthen classroom norms and support more productive use of available practice time. As learning platforms become increasingly integrated into classroom instruction, understanding and addressing how students use (and underuse) practice time is fundamental to realizing the learning benefits of these systems.

\section{Acknowledgments}
This work was supported by the Learning Engineering Virtual Institute. The opinions, findings, and conclusions expressed in this material are those of the authors and do not reflect the views of the Institute. The authors also thank WestEd and the ASSISTments Foundation for their commitment to open data and open science, which made this secondary analysis possible.

\bibliographystyle{ACM-Reference-Format}
\bibliography{main}

\end{document}